\documentclass[
reprint,
 amsmath,amssymb,
 aps,
]{revtex4-2}

\usepackage{graphicx}
\usepackage{dcolumn}
\usepackage{bm}
\usepackage{xcolor}
\usepackage{ulem}

\begin{document}

\title{Sample-dependent Dirac point gap in MnBi$_2$Te$_4$ and its response to the applied surface charge: a combined photoemission and ab initio study}

\author{A.M. Shikin\textsuperscript{1,†}}
\email{ashikin@inbox.ru}

\author{D.A. Estyunin\textsuperscript{1}}
\thanks{These two authors contributed equally}

\author{N.L. Zaitsev\textsuperscript{2}}
\author{D. Glazkova\textsuperscript{1}}
\author{I.I. Klimovskikh\textsuperscript{1}}
\author{S. Filnov\textsuperscript{1}}
\author{A.G. Rybkin\textsuperscript{1}}
\author{E. F. Schwier\textsuperscript{3}}
\author{S. Kumar\textsuperscript{3}}
\author{A. Kimura\textsuperscript{4}}
\author{N. Mamedov\textsuperscript{5}}
\author{Z. Aliev\textsuperscript{5,6}}
\author{M.B. Babanly\textsuperscript{7}}
\author{K. Kokh\textsuperscript{1,8,10}}
\author{O.E. Tereshchenko\textsuperscript{1,9,10}}
\author{M.M. Otrokov\textsuperscript{11,12}}
\author{E. V. Chulkov\textsuperscript{1,13,14}}
\author{K.A. Zvezdin\textsuperscript{15,16}}
\author{A.K. Zvezdin\textsuperscript{15,16}}

\affiliation{
\textsuperscript{1} Saint Petersburg State University, 198504 Saint Petersburg, Russia\\
\textsuperscript{2} Institute of Molecule and Crystal Physics, Ufa Research Center of Russian Academy of Sciences, 450075, Ufa, Russia\\
\textsuperscript{3} Hiroshima Synchrotron Radiation Center, Hiroshima University, Hiroshima, Japan\\
\textsuperscript{4} Department of Physical Sciences, Graduate School of Science, Hiroshima University, Hiroshima, Japan\\
\textsuperscript{5} Institute of Physics, ANAS, AZ1143 Baku, Azerbaijan\\
\textsuperscript{6} Azerbaijan State Oil and Industry University, AZ1010 Baku, Azerbaijan\\
\textsuperscript{7} Institute of Catalysis and Inorganic Chemistry, ANAS, AZ1143 Baku, Azerbaijan\\ 
\textsuperscript{8} V.S. Sobolev Institute of Geology and Mineralogy, Novosibirsk, 630090, Russia\\
\textsuperscript{9} A.V. Rzhanov Institute of Semiconductor Physics, Novosibirsk, 630090, Russia\\
\textsuperscript{10} Novosibirsk State University, Novosibirsk, 630090, Russia\\
\textsuperscript{11} Centro de F\'isica de Materiales, Facultad de Ciencias Químicas, UPV/EHU, Apdo. 1072, 20080 San Sebasti\'an, Spain\\
\textsuperscript{12} IKERBASQUE, Basque Foundation for Science, E-48011 Bilbao, Basque Country, Spain\\
\textsuperscript{13} Donostia International Physics Center (DIPC), 20018 Donostia-San Sebastián, Basque Country, Spain\\
\textsuperscript{14} Departamento de Pol\'imeros y Materiales Avanzados: F\'isica, Qu\'imica y Tecnolog\'ia, Facultad de Ciencias Qu\'imicas, Universidad del Pa\'is Vasco UPV/EHU, 20080 San Sebasti\'an/Donostia, Basque Country, Spain.\\
\textsuperscript{15} A.M. Prokhorov General Physics Institute, Russian Academy of Sciences, Moscow, 119991, Russia\\
\textsuperscript{16} Russian Quantum Center, Skolkovo, 143025, Russia}
\date{\today}%

\begin{abstract}

Recently discovered intrinsic antiferromagnetic topological insulator MnBi$_2$Te$_4$ presents an  exciting platform for realization of the quantum anomalous Hall effect and a number of related phenomena at elevated temperatures. An important characteristic making this material attractive for applications is its predicted large magnetic gap at the Dirac point (DP). However, while the early experimental measurements reported on large DP gaps, a number of recent studies claimed to observe a gapless dispersion of the MnBi$_2$Te$_4$ Dirac cone. Here, using micro($\mu$)-laser angle-resolved photoemission spectroscopy, we study the electronic structure of 15 different MnBi$_2$Te$_4$ samples, grown by two different chemists groups. Based on the careful energy distribution curves analysis, the DP gaps between 15 and 65 meV are observed, as measured below the N\'eel temperature at about 10-16 K. At that, roughly half of the studied samples show the DP gap of about 30 meV, while for a quarter of the samples the gaps are in the 50 to 60 meV range. Summarizing the results of both our and other groups, in the currently available MnBi$_2$Te$_4$ samples  the DP gap can acquire an arbitrary value between a few and several tens of meV. Further, based on the density functional theory, we discuss a possible factor that might contribute to the reduction of the DP gap size, which is the excess surface charge that can appear due to various defects in surface region. We demonstrate that the DP gap is influenced by the applied surface charge and even can be closed, which can be taken advantage of to tune the MnBi$_2$Te$_4$ DP gap size. 

\end{abstract}

\maketitle

\section*{Introduction}

The realization of the quantized anomalous Hall (QAH) and magnetoelectric effects \cite{Qi2008,Qi2011,Chang2013, Liu2016,Tokura2019,Chang2016,Wang2015,Essin2009} in magnetic topological insulators (TI) requires opening the energy gap at the Dirac point (DP) due to the out-of-plane magnetization. The size of this gap defines capability  of a magnetic TI to show the above-mentioned effects at high temperatures, which opens a quest for materials with  large DP gaps. 

Recently, the first intrinsic antiferromagnetic (AFM) TI MnBi$_2$Te$_4$ (MBT) has been discovered and extensively investigated \cite{Otrokov.2dmat2017,Otrokov.jetpl2017,Eremeev.jac2017,Otrokov.prl2019,Otrokov2019, Li2019, Zhang2019, Aliev2019, Yan_PRM2019, Zeugner2019,Li.prl2020,Deng2020,Liu_NatMat_2020,Petrov.arxiv2020,Ge2020,Xu.prb2021,Perez-Piskunow.prl2021,Wang.inn2021}. This layered compound consists of the septuple layer (SL) blocks with a stacking sequence of Te-Bi-Te-Mn-Te-Bi-Te  \cite{Aliev2019, Zeugner2019}. The neighboring blocks are separated by van der Waals (vdW) spacings. Within each SL, Mn atoms are ordered ferromagnetically, while the adjacent SLs are coupled  in an AFM fashion \cite{Yan_PRM2019, Li.prl2020}, as shown in Fig.1a. Density functional theory (DFT) calculations predict the MBT(0001) surface to exhibit a giant DP gap up to about 90 meV, opened below the N\'eel temperature  $T_N$ = 25 K \cite{Otrokov2019,Zhang2019,Li2019}. In the two-dimensional limit, MBT has been theoretically predicted to show the quantized Hall effect both with and without external magnetic field as well as the axion insulator state \cite{Li2019,Otrokov.prl2019}, which has later been confirmed experimentally \cite{Deng2020,Liu_NatMat_2020,Ge2020}. Moreover, MBT has given rise to a family of the alike compounds such as (MnBi$_2$Te$_4$)(Bi$_2$Te$_3$)$_{n}$ \cite{Aliev2019,Wu2019,Hu2020,Klimovskikh2020,Jahangirli2019,Vidal.prx2019}, MnBi$_{2-x}$Sb$_x$Te$_4$ \cite{Chen_Nat_Com_2019,Yan_PRB_2019, Ko.prb2020, Abdullaev.ftt2021}, (MnSb$_2$Te$_4$)(Sb$_2$Te$_3$)$_n$ \cite{Eremeev.jac2017,Yan_PRB_2019,wimmer2021mnrich,Eremeev.jpcl2021,huan2021multiple, Huan.apl2021}. Recently, the QAH effect has been reported for  MnBi$_{10}$Te$_{16}$ \cite{Deng2021}.

In spite of the theoretical predictions of a large DP gap at the MBT(0001) surface, from the experimental side there exist significant discrepancies concerning the measured DP gap size, as the angle-resolved photoemission spectroscopy (ARPES) studies report rather contradicting results. The first ARPES studies, performed with synchrotron radiation \cite{Otrokov2019, Lee2019, Vidal.prb2019}, indeed revealed the presence of the Dirac cone at the (0001) surface featuring a gap of about 70-100 meV. Laser-based ARPES provided further insights into the MBT surface electronic structure \cite{Otrokov2019,Hao2019,Li2019.prx,Chen2019,Swatek2019,Estyunin2020,Shikin2020a,Nevola2020,Yan.arxiv2021}. Using this technique, several groups reported \cite{Hao2019,Chen2019,Li2019.prx,Swatek2019,Nevola2020,Yan.arxiv2021} a completely opposite behavior of the MBT Dirac cone, claiming its dispersion to be gapless. Noteworthy, the recent laser-ARPES study reporting both large ($\sim 50$ meV) and small (but finite; $\leq 20$ meV) DP gaps \cite{Shikin2020a}.

Concerning possible reasons of the gapless behavior of the  topological surface state (TSS), it can be achieved in a DFT calculation by forcing the system to the in-plane A- or G-type AFM  orders \cite{Hao2019,Swatek2019}. However, both cases disagree with direct magnetic measurements by neutron scattering and SQUID which reveal only the out-of-plane A-type antiferromagnetism \cite{Otrokov2019,Yan_PRM2019,Zeugner2019,Li.prl2020}. Moreover, such magnetic orderings appear to be significantly less favorable in energy than the A-type AFM order \cite{Eremeev.jac2017,Hao2019}. Therefore, one can expect that the emergence of a gapless or a small-DP-gap TSS in MBT is due to a deviation in magnetic ordering or/and crystal structure of the topmost SL from the ideal one(s). For instance, as proposed in Ref. \cite{Hao2019} a magnetically disordered (i.e. paramagnetic) layer might be formed near the MBT surface, because of which TSS would experience no exchange splitting. However, this idea contradicts to the recent reports \cite{Estyunin2020,Nevola2020}, which observe a temperature-dependent Rashba-like state in MBT. This state might have a surface localization \cite{Bahramy2012}, and it clearly demonstrates the band gap opening at the Kramers point below  $T_N$  both  in the sizable \cite{Estyunin2020} and vanishing \cite{Nevola2020} DP gap cases. Moreover, recent magnetic force microscopy \cite{Sass2020} and X-ray magnetic circular dichroism measurements \cite{Otrokov2019,Shikin2020a} do not corroborate the magnetism  change at the MBT surface. These observations  indicate the retention of the FM out-of-plane magnetic order in the topmost SL  MBT(0001), regardless of the difference in the measured DP gaps. 

On the structural side, STM investigations do not show any serious degradation or reconstruction of the MBT(0001) surface, at least in the ultra-high vacuum conditions \cite{Yuan2020, Yan_PRM2019, Liang.prb2020, Huang.prm2020, Hou2020}. However, on the atomic level, the surface shows a significant amount of the point defects caused mainly by the Mn-Bi intermixing leading to the appearance of the Mn$_\text{Bi}$ (i.e. Mn replaces Bi) \cite{Yuan2020, Yan_PRM2019, Liang.prb2020, Huang.prm2020, Hou2020} and Bi$_\text{Mn}$ atoms \cite{Huang.prm2020}. Besides, the Bi$_\text{Te}$ antisites are also presented, although in a much smaller amount \cite{Yuan2020, Yan_PRM2019, Liang.prb2020, Huang.prm2020, Hou2020}. Although not seen in the STM topographs, Mn vacancies ($V_\text{Mn}$) could also be present as suggested by the results of the bulk XRD \cite{Zeugner2019} and TEM \cite{Lee2019} measurements. The charge states of these defects were reported to be Mn$^-_\text{Bi}$, Bi$^+_\text{Mn}$, Bi$^-_\text{Te}$, and $V^{2+}_\text{Mn}$ \cite{Du.advfunmat2020}. The concentration of these defects may experience certain variations inside the sample. Besides, there may be other reasons affecting the surface quality and properties such as adatoms, cracks, steps between terraces, etc. Finally, the effects of photostimulated  adsorption and ionisation are pronounced during ARPES measurements of TIs \cite{Frantzeskakis2017}. All the mentioned factors may result in some uncompensated charge, that can affect the electronic  structure of the surface layers, however the influence of this charge on the TSS has not been studied so far.

In this paper we present a combined laser-ARPES and DFT study of the MBT(0001) surface electronic structure with a focus on the DP gap. On the experimental side, we present the band structures measured with ARPES for 15 different MBT samples grown by two different chemists groups. In the AFM phase at about 10-16 K, various sizes of the DP gaps have been observed ranging from relatively small ($\sim15$~meV) to large ones ($\sim65$~meV), with half of the samples showing gaps of about 30 meV. In conjunction with the experimental evidence available from literature, our results indicate that in the currently available MBT samples the DP gap can acquire an arbitrary value between a few and several tens of meV. On the theoretical side, in order to take the influence of possible uncompensated surface charge into account we introduce this charge as an effective variable parameter in our DFT calculations and study changes in the TSS. The applied electric field not only shifts the TSS within the bulk band gap (like in non-magnetic TIs (Bi,Sb)$_2$(Te,Se)$_3$ \cite{Yazyev.prl2010,Menshchikova2013}), but also affects the DP gap size, which significantly decreases in the case of negative charge. This is accompanied by a redistribution of the TSS real space density within three topmost SL and alteration of atoms magnetic moments that can be possible reasons for DP gap size change. These results show that an uncompensated charge on the MBT surface might contribute to the reduction of the DP gap size observed in some MBT samples. From the practical point of view, it can be used to tune the DP gap size, what could enable switching on/off the QAH or axion insulator state in the MBT thin films.

\section*{Results and discussion}

\begin{figure}
\includegraphics[width=0.5\textwidth]{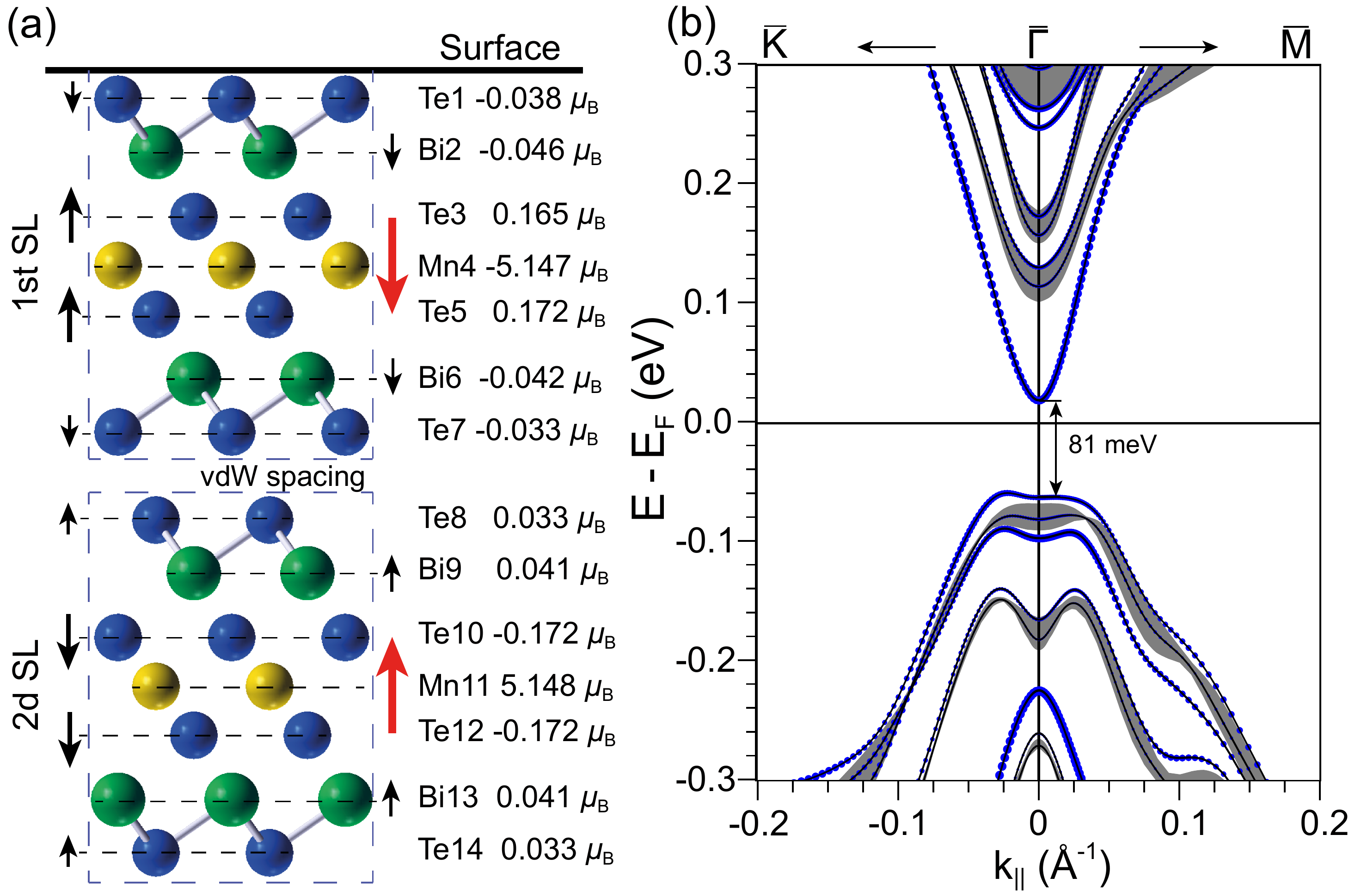}
 \caption{(a) - Schematic crystalline structure of the first two SLs in MBT with marked direction and value of magnetic moment on each atom. (b) - Calculated electronic structure of surface states in 6 SL MBT slab without applying external charge. The Dirac gap size is 81 meV. The size of circles reflects the degree of localization of the corresponding wave function within the first SL. The gray areas mark the bulk bands projected onto the surface Brillouin zone.}
 \label{fig_cryst_str}
\end{figure}

Fig. 1a schematically presents two SLs of the near-surface MBT crystal structure. Mutual orientation of the local and induced magnetic moments on atoms is shown by arrows of different sizes, while the absolute values of the magnetic moments are given next to the structure sketch on the right. Both the directions and values of the magnetic moments are taken from ab-initio calculations (see methods section). Within a single SL, the Mn atoms induce the moments of the same direction on the Bi and vdW Te layer (e.g. Te1 and Te7), and the moments of the opposite direction on the inner Te atoms (e.g. Te3 and Te5). Due to the interlayer AFM ordering, magnetic moments of the atoms in the second SL are antiparallel to those of the corresponding atoms in the first SL. Fig. 1b shows DFT-calculated surface and bulk projected bands of MBT(0001). One can see that both parts of the DC are located inside the bulk band gap at zero applied electric field.

\begin{figure*}
\includegraphics[width=0.9\textwidth]{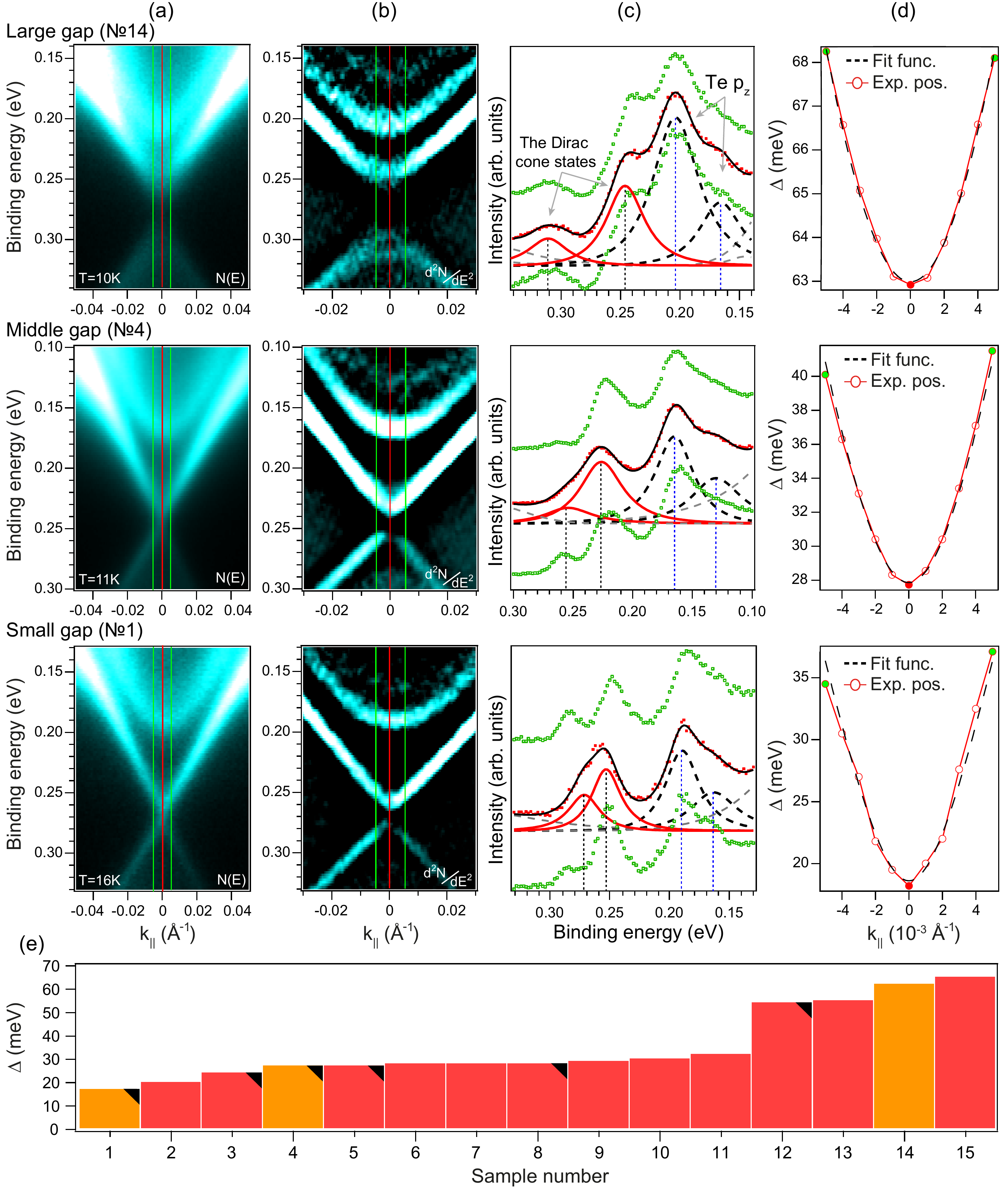}
 \caption{Column (a) - ARPES dispersion relations measured at $h\nu=6.3$~eV with s-polarized LR below MnBi$_2$Te$_4$ (10 - 16 K) for MBT samples with large, medium and small Dirac gaps and their the second derivatives $d^2N/dE^2$ in column (b). Column (c) - EDCs cut at the $\Gamma$-point  ($k_\|=0$~\AA$^{-1}$) - red curves and some $\pm k_\|$ green curves. Position of the cuts are marked with lines of corresponding color in columns (a,b). Spectral components decomposition are shown for EDCs cut at the $\Gamma$-point. Red peaks correspond to the Dirac cone states, black peaks - to the Te $p_z$ exchange-split states.  Additional peaks that are out of interest are gray. Column (d) dependence of the splitting between the top and bottom Dirac cone states on wave vector ($k_\|$). Dots (connected with lines) show experimental values of splitting, black curve is fitting function of the gapped cone (see text).  (e) -  the DP gap sizes of all 15 MBT samples presented in this paper. Orange (red) bars correspond to the samples whose spectra are presented in Fig.2 (Suppl.Inf.Fig1S-3S). Samples from the Novosibirsk group are marked by black triangles, while the rest of the samples have been grown in Baku.} 
 \label{fig_ARPES1}
\end{figure*}

Fig.2 column (a)  shows the experimental ARPES dispersion relations presented in the energy region, which includes the TSS together with the bulk states in the valence and conduction bands (VB and CB, respectively). The measurements were performed for different MBT samples. Column (b) shows the second derivatives of the N(E) shown in (a) for better visualization. The dispersion relations were measured using laser radiation (LR) with photon energy of 6.3 eV below T$_{\rm N}$ (10-16 K). The position of the sample was adjusted to be right at $k_{\|y}=0 $~\AA$^{-1}$ by pre-measured $E(k_{\|x}, k_{\|y})$ mappings and then dispersion relations were measured along $k_{\|x}$ (shown as $k_\|$ in the figures).

Column (c) shows the energy distribution curves (EDCs) at the $\Gamma$-point ($k_{\|}=0 $~\AA$^{-1}$; red  dots) and in its vicinity for $\pm k_\|$ (green color). Positions of profiles are marked in the columns (a,b) by vertical lines of corresponding colors. Decomposition of the EDCs at the $\Gamma$-point into the spectral components is also presented. One can distinguish the DC components (two red peaks for top and bottom parts of the cone) and the exchange-split Te $p_z$ bulk CB states (two black dashed peaks), which are clearly visible in column (b) panels (see Ref.~\cite{Estyunin2020} for more details on the EDC analysis). The onset of the splitting of the Te $p_z$  states can serve as an indicator of the magnetic transition in  MBT. Column (d) shows dependence of the energy separation between the top and bottom parts of the DC on $k_\|$. The dots filled with red and green colors mark the values taken from the corresponding EDCs in column (c). The experimental points are fitted with the model curve for the gapped cone $E \sim (\alpha^2k^2+\Delta^2)^{1/2}$, where $\Delta$ -- is the DP gap size. This approach allows reducing the estimation error ($\sim5$~meV). The data acquired on all of the studied samples (in Fig.2 and in Figs.1S-3S) were treated equally and are presented in a similar form as described above.

Going from the top to the bottom, the DP gap sizes (i.e. the separations between the red peaks in Fig.2c) are $\Delta E_{DG} = 63$~meV, 28~meV, 18~meV. Here one should keep in mind the experimental error which, however, does not exceed $\sim5$~meV. Other series of the experiments (see Suppl. Inf. Figs.1S-3S) demonstrate similar variability  of the DP gap sizes for different samples. The values of the DP gap  measured from all 15 MBT samples are presented in Fig.2e. Here we observe results practically filling the whole energy interval between 15 and 65 meV. Moreover, one can see that for half of the samples the DP gaps tend to be in the 28-33 meV range (i.e. "middle gap"). Importantly, the results obtained do not depend on the sample origin, as the samples from both groups basically show the whole spectrum of the gaps, from small to large.  Together with other experimental reports available from literature, our results indicate that in the currently available MBT samples the DP gap can acquire an arbitrary value between a few and several tens of meV. 

As far as the bulk states are concerned, from the top  to the bottom line the exchange splitting between the CB Te $p_z$ states (i.e. the separation between black peaks Fig.2c) are $E_{ex}=38$~meV (T = 10 K), 36~meV (T = 11 K) and 27~meV (T = 16 K). Including the temperature dependence for the Te $p_z$ states as $E_{ex} = E_0\cdot(1-T/T_0)^{1/2}$, the splitting onset temperature $T_0$ turns out to be approximately equal to $T_{\rm N}$ \cite{Estyunin2020}, while at $T=0$~K  a value of about $\sim45$~meV is achieved for saturated splitting $E_0$ in all three cases.  Thus, while the DP gap size varies significantly from sample to sample, the size of the (saturated) exchange splitting of the Te $p_z$ states remains almost the same.

\begin{figure}
\includegraphics[width=0.5\textwidth]{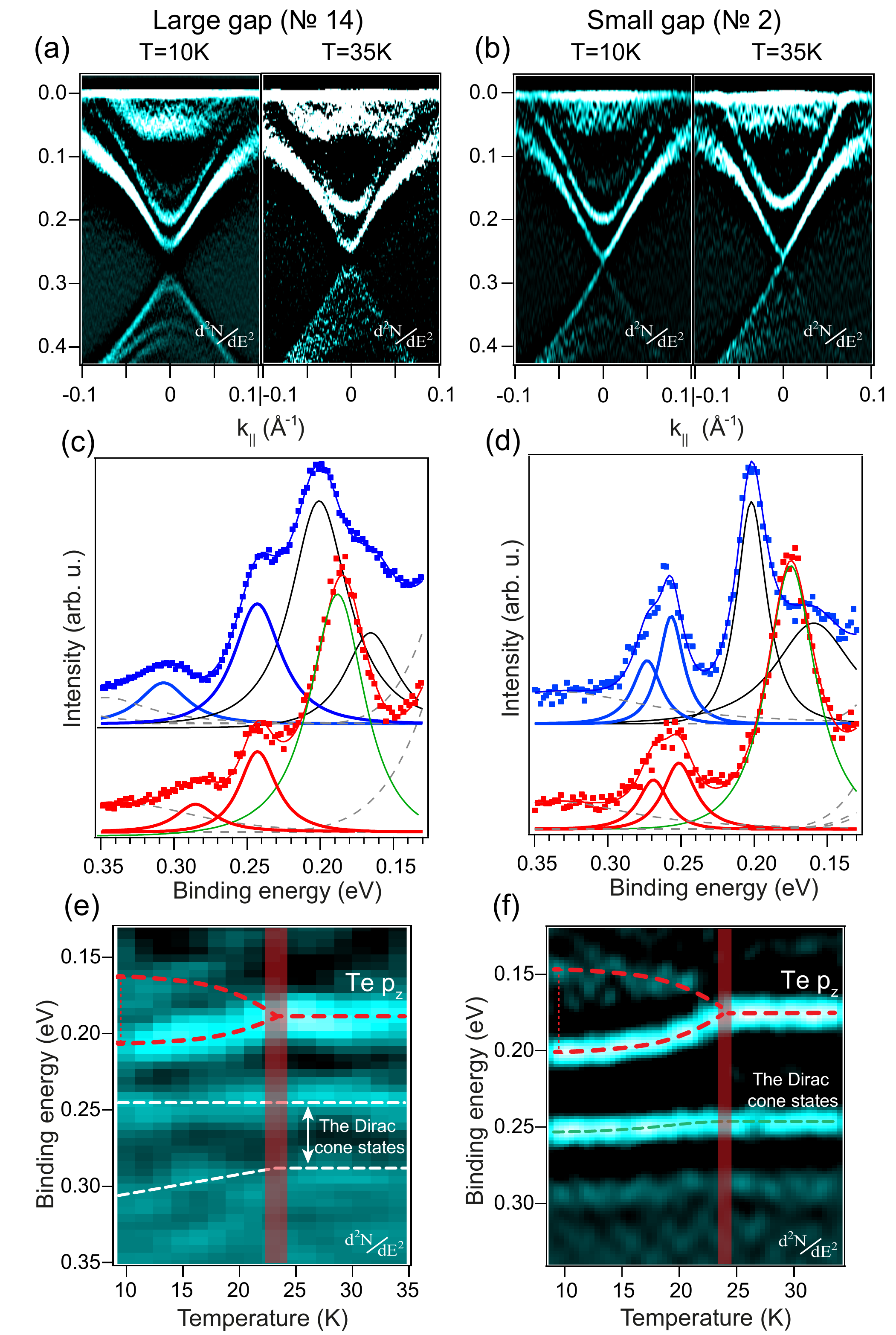}
 \caption{(a,b) - ARPES dispersion relations measured for MBT below (10 K) and above T$_{\rm N}$ (35 K) for samples with the different Dirac gap values in the $d^2N/dE^2$ presentation for better visualization of the features. (c,d) - EDCs at 10 K (blue curves) and 35 K (red curves) cut at the $\Gamma$-point ($k_\|=0$~\AA$^{-1}$) with decomposition to spectral components: solid line peaks of corresponding color - the Dirac cone states, black and green peaks - Te $p_z$ states, gray peaks - VB and CB states. (e,f) - the modifications (as $d^2N/dE^2$) of the states at the $\Gamma$-point under permanent growth of temperature from 10 K to 35 K (step $\Delta$T = 1.5 K and 0.5 K). Red and white (blue) curves are guide to the eye temperature variation of Te $p_z$ and the Dirac cone states, respectively. Vertical half-transparent line marks onset temperature of Te $p_z$ splitting. }
 \label{fig_temp}
\end{figure}

Fig. 3 shows the dispersion relations (as $d^2N/dE^2$ for better visualization) measured for two different samples below and above  $T_{\rm N}$ as well as the temperature dependence of the states at the $\Gamma$-point in the 10-35~K temperature interval. One can see that the spectra for large (a) and small (b) DP  gap cases reveal practically similar changes  upon going from 10 K to 35 K: modification of the Rashba-like state at the Fermi level, merging of the Te $p_z$ states (at a BE of $0.16-0.20$~eV) and drop of the intensity in the DP region. The last two modifications are clearly visible in panels (c,d) which show the spectral components decompositions of the EDCs  at the $\Gamma$-point for both 10 K (blue) and 35 K (red). Namely,  both in (c) and (d), the two black peaks of the Te $p_z$ states that are clearly separated at 10 K merge at their center of mass to form a single (green) peak at a BE$\sim0.18$~eV above the N\'eel temperature. The intensity drop is visible for the Dirac cone components (blue and red peaks) and the ratio between the peaks intensities at 10 K and 35 K is about 1.6 both for (c) and (d). From the temperature dependence in panels (e) and (f), we estimate the onset temperature of the Te $p_z$ states splitting in agreement with  $T_{\rm N}$. This onset temperature is  estimated  to be about 24 K (23.5 K) for the small (large) DP gap case as marked by the semi-transparent vertical stripes. These values can be considered as practically equal within experimental error ($\sim3$ K).  This splitting is a good fingerprint of the emerging bulk magnetism of MBT \cite{Otrokov2019,Estyunin2020}. Thus, the magnetic properties for both  samples shown in Fig. 3 seem to be similar despite the different DP gap sizes.

Similarly to  the previous findings \cite{Lee2019,Vidal.prb2019,Estyunin2020,Shikin2020a}, the DP gaps remain open above $T_{\rm N}$ in both cases presented in Fig.3. However, the large gap  gets smaller in the paramagentic phase. Indeed, as seen in Fig.3(c,e), under heating the peak of the lower part of the DC shifts to lower BE by about 25 meV. The upper part of the DC remains unshifted.  As a result, the DP gap size is reduced from $\sim65$~meV at 10 K to  $\sim40$~meV at  35 K (c). This behavior is schematically (i.e. without proper fitting) shown in panel (e). In case of the small gap both components of the DC move jointly under heating up to  $T_{\rm N}$ (above  $T_{\rm N}$ their positions are constant) and so the DP gap is unchanged. 

As it has been said above, structural  studies of the MBT bulk \cite{Zeugner2019, Lee2019} and surface \cite{Yuan2020, Yan_PRM2019, Liang.prb2020, Huang.prm2020, Hou2020}  reveal presence of various point defects. These are the Mn$^-_\text{Bi}$ and Bi$^+_\text{Mn}$ substitutions, Bi$^-_\text{Te}$ antisites, and $V^{2+}_\text{Mn}$ vacancies \cite{Du.advfunmat2020}. On average, MBT bulk is likely charge neutral, i.e. charged point defects compensate each other within some sample's volume. If this volume is split, for example, during cleavage process, an uncompensated charge can appear at the created surface. Additionally, plenty of other defects at the sample surface can exist during APRES experiment: adatoms, cracks, steps with one or more SLs height and other. Moreover, effects of photostimulated adsorption and ionisation in an ARPES experiment are important in TIs \cite{Frantzeskakis2017}. Therefore, some charge excess, either positive or negative, can appear on the surface for many various reasons. Let us now study their influence on TSS via application of the external charge to the surface of MBT. 

\begin{figure*}
\includegraphics[width=\textwidth]{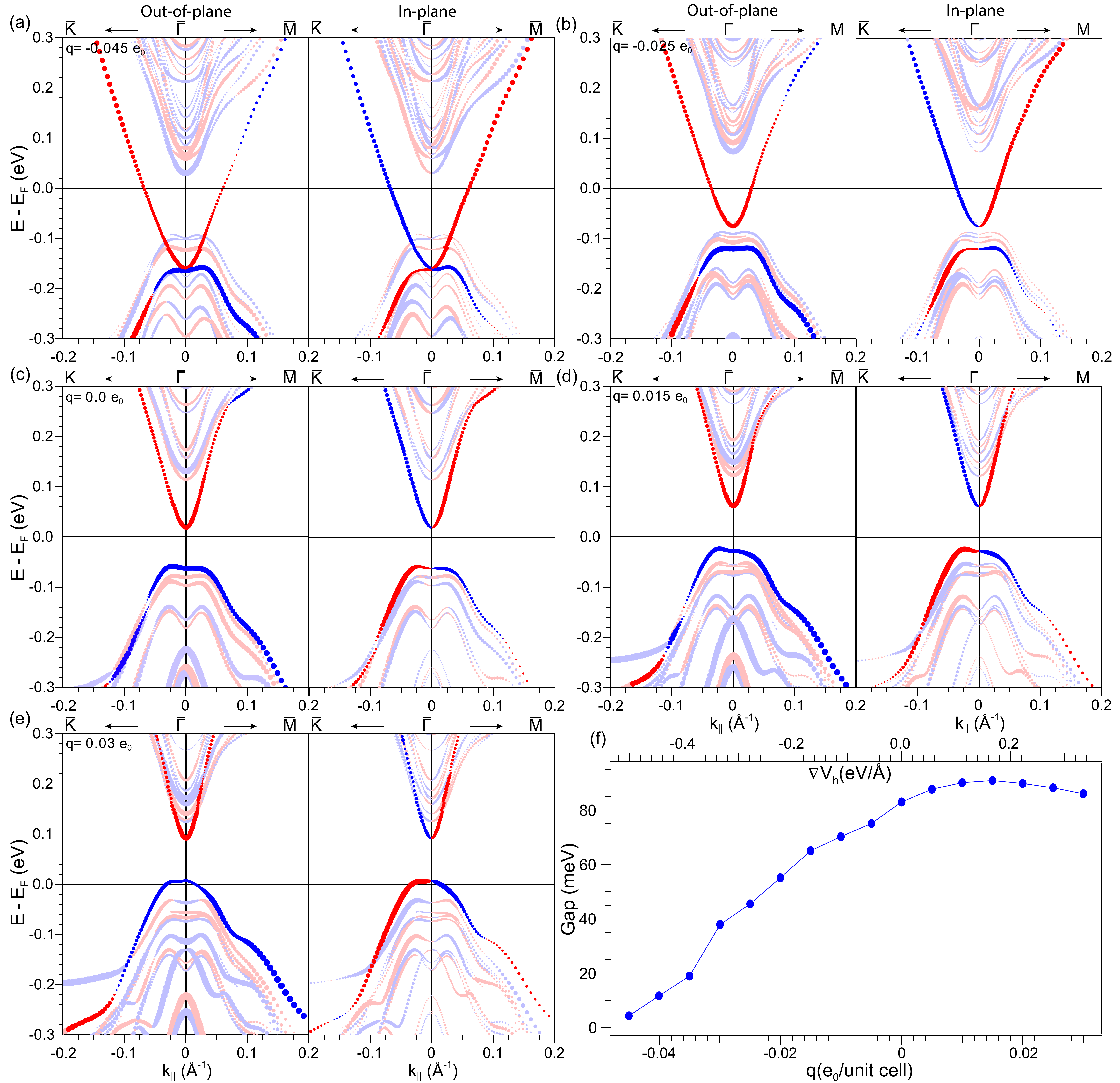}
 \caption{(a-e) - Spin-resolved calculated dispersion relations with various applied charges to the sample surface. Left and right panels show out-of-plane (+S$_z$ - red and -S$_z$ - blue) and in-plane (+S$_{x,y}$ - red and -S$_{x,y}$ - blue) spin components, size of circles matches the value of spin-vector projection. The DC state is roughly highlighted with brighter colors. (f) - The change of the Dirac gap value plotted depending on the surface charge (bottom axis) and the surface potential gradient (top axis).}
 \label{fig_Calc}
\end{figure*}
 
To do this, we have carried out  DFT calculations with an applied charge  varied in the range from -0.045~e$_0$ to +0.03~e$_0$.  Fig. 4c  shows the calculated band structure of the MBT film without applied charge (also see Fig.1b for bulk projected bands). The gapped TSS is highlighted and both the out-of-plane and in-plane spin components are shown in the right and left subpanels, respectively. When a positive (negative) charge is applied, one can notice an upward (downward)  shift of the bands as compared to the  unperturbed bands  due to the band bending. At that,  the TSS shows much larger shifts as compared to the bulk-like states of the film.  For instance, at about $-0.025$ e$_0$ (Fig.4b) both the top and  bottom parts of the gapped DC  start to overlap with the bulk VB states  around $k_\|=0$~\AA$^{-1}$ or, alternatively, shift towards the bulk band gap center at about $0.01 - 0.015$ e$_0$ (Fig.4d). This is similar to what has been reported for non-magnetic TIs (Bi,Sb)$_2$(Te,Se)$_3$ under the electric field \cite{Yazyev.prl2010,Menshchikova2013}. It should also be noted that irrespectively of the charge sign and value, in vicinity of the $\Gamma$-point, the DC shows  a “hedgehog”  spin texture (i.e., the in-plane component is helical, while the  out-of-plane one is non zero and opposite for top and bottom parts of the cone). 

\begin{figure*}
\includegraphics[width=\textwidth]{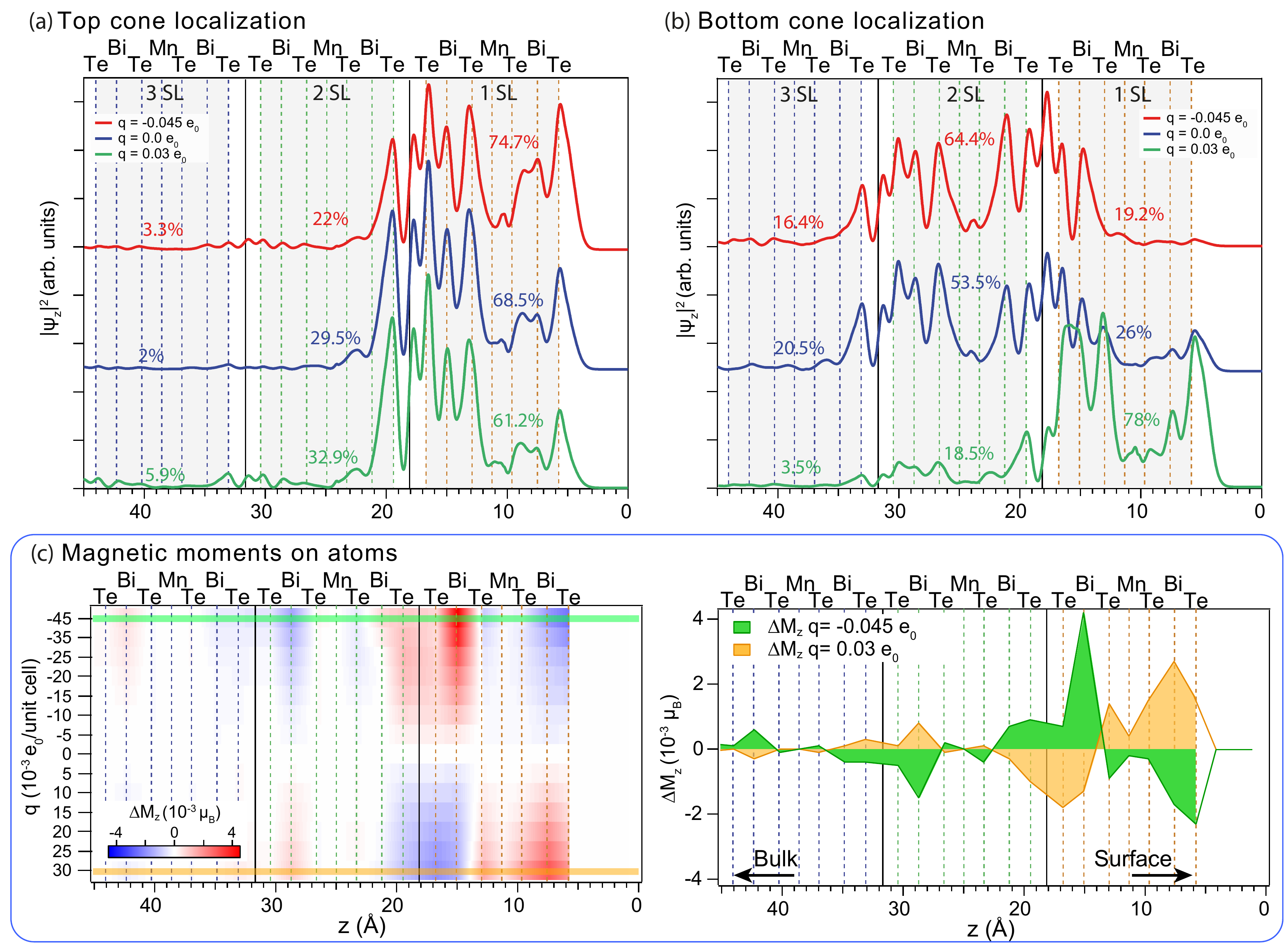}
 \caption{Spatial density distribution along $z$ axis of the top (a) and bottom (b) parts of the DC taken at the $\Gamma$-point for various applied charges q=-0.045 e$_0$ (red), 0 e$_0$ (blue) and +0.03 e$_0$ (green). Weight of the upper (lower) part of TSS in each SL is shown by numbers of corresponding color.  Left part of (c) - spatial distribution of magnetic moments on MBT atoms for various applied charges presented as ($\Delta M(z,q)= M(z,q)-M(z,0)$). Color represents value of $\Delta M(z,q)$, horizontal lines mark positions of profiles shown in right  part of (c). }
 \label{fig_Calc}
\end{figure*}

Let us now focus on the DP gap response to the applied charge. The dependence of the DP gap size on the surface charge and the corresponding gradient of the surface potential is presented in Fig.4f.  One can see a strong  decrease of the DP gap with an applied negative charge: it drops from 81 meV to almost zero ($\sim5$~meV) when the charge magnitude is changed from 0 to -0.045 e$_0$. For the positive charge, the DC gap first slightly increases up to 89 meV at +0.015 e$_0$ and then goes down approximately to its initial value  at +0.03 e$_0$.

The above described changes of the DP gap size are accompanied by certain changes in spatial distribution (along z-axis) of the TSS density under applied surface charge as shown in Figs.5a,b. In the charge neutral case, about 70 \% (30 \%) of the real space weight of the upper part of the gapped TSS is located in the first (second) SL block. In turn, 50 \% of the lower part of the state is localized in the second SL, while the first and the third septuples host the rest in almost equal shares. When a negative excess charge is applied to the surface, the upper part of the TSS increases its localization in the surface SL, up to about 75 \% for a charge value of -0.045 $e_0$. In contrast, at this value of the applied charge, the lower part of the TSS gets further shifted in the subsurface SL block, which now hosts about 65 \% of the state.

Such a redistribution is likely to affect the DP gap size. Indeed, since the two SLs are coupled antiferromagnetically, they exert the exchange fields of the opposite signs on the TSS, which should lead to the decrease of the DP gap shown in Fig. 4. As far as the positive excess charge is concerned, it slightly moves the upper part of the TSS into the second and third SLs, without significantly changing its distribution even up to a charge value of +0.03~$e_0$. In turn, the lower part of the TSS gets strongly pushed into the surface SL block. Obviously, the largest DP gap values should be achieved when both the upper and lower parts of the gapped TSS are mostly located in the same SL or in the SLs that are next nearest to each other, as in this case they experience an exchange field of the same sign. This is exactly what happens up to +0.015~$e_0$, i.e. the gap is increasing. However, since with the positive charge increase the upper part slightly reduces its net weight in the "first + third" SLs in the favor of the second SL, while the lower part abandons the third SL, this leads to a certain decrease of the DP gap value seen in Fig. 4f. 

We also notice some modification of magnetic moments, $\Delta$M(z,q), of MBT atoms under the applied charge (Fig.5c). It should be noted that while the Mn local moment magnitude ($\sim5.15~\mu_B$) significantly exceeds those of Te ($\sim0.17$ and $\sim0.03~\mu_B$) and Bi ($\sim0.04~\mu_B$) [Fig.1a], it is the latter atoms that bear most of the TSS weight (Fig.5 a,b). Therefore, magnetic moments of Te and Bi might play an important role in the DP gap size. In Fig.5c one can see that positive and negative surface charges induce the variation $\Delta$M(z,q) of opposite sign for each atom, thus demonstrating a magnetoelectric effect. The amplitudes of variation depend both on the atomic sort and the depth of the layer location. While the change of the Mn magnetic moment relative to its absolute value is negligible, the Te and Bi magnetic moments change significantly especially in the 1st SL. The largest value of $\Delta$M(z,q) is $\sim0.004~\mu_B$, which is about 10 \% of the initial moments of the Bi and vdW Te atoms (e.g. Te1 and Te7). 

Although being small, such an M(z,q) variation can nevertheless affect the TSS and may be an additional factor that influences the DP gap size. Indeed, e.g. in the case of graphene, inducing only $\sim0.002~\mu_B$ per carbon atom (as calculated within a Wigner-Seitz sphere) is enough to achieve an exchange-splitting of 175 meV \cite{Rybkin.nl2018}. Although the electronic structures of graphene and MBT are different (and therefore such comparisons should be taken with caution), the latter example illustrates that small magnetic moment does not necessarily mean negligible exchange splitting.

Going back to the excess surface charge effect on the MBT TSS, we speculate that it can be used to tune the DP gap size, e.g., to enlarge the gap in the samples where it is strongly reduced. On the other hand, upon applying a sufficiently large negative charge the DP gap is expected to shrink, while reducing the negative charge magnitude will open it again. This can potentially be used to switch off and on the QAH or axion insulator states in the MBT thin films with even or odd number of SLs, respectively. Closing the DP gap would render both systems two-dimensional metals, while its reopening would restore the respective QAH and axion insulator phases.

\section*{Conclusion}

In the present work, we experimentally observed a variety of the DP gap sizes measured for 15 different samples of AFM TI MnBi$_2$Te$_4$ (MBT) grown by two different chemists groups. In the AFM phase at about 10-16 K, we have measured DP gaps ranging from relatively small ($\sim15$~meV) to large ones ($\sim65$~meV), with half of the samples showing gaps of about 30 meV. The results obtained on the samples from both groups are consistent with each other, i.e., they show the whole spectrum of the gaps, from small to large. In conjunction with the experimental evidence available from literature, our results indicate that in the currently available MBT samples the DP gap can acquire an arbitrary value between a few and several tens of meV. 

Our DFT calculations show that an uncompensated surface charge, which can possibly emerge from structural defects at the MBT surface, might contribute to the reduction of the DP gap size observed in some MBT samples. From the  practical  point  of  view,  it  can  be  used  to  tune  the DP  gap  size,  what  could  enable  switching  on/off the QAH  or  axion  insulator  state  in  the  MBT  thin  films.

\section*{Methods}

High-quality MnBi$_2$Te$_4$ single crystals were synthesized using the vertical Bridgman method and characterized by X-ray diffraction at the Azerbaijan State Oil and Industry University and Institute of Physics ANAS and Novosibirsk State University.

The measurements of the ARPES dispersion maps were carried at the $\mu$-Laser ARPES system at HiSOR (Hiroshima, Japan) with improved angle and energy resolution and a high space resolution of the laser beam (spot diameter around 5 $\mu$m) using a Scienta R4000 analyzer with an incidence angle of the LR of $50^\circ$ relative to the surface normal. For the experiment we used photons with $h\nu=6.3$~eV.

Clean surfaces of the samples were obtained by a cleavage in ultrahigh vacuum. The base pressure during all photoemission experiments was better that $1\times 10^{-11}$ mbar.

The electronic structure calculations were done by using OpenMX code, providing the fully relativistic DFT (density functional theory) implementation with localized pseudoatomic orbitals \cite{Ozaki2003,Ozaki2004,Ozaki2005} and the norm-conserving pseudopotential \cite{Troullier1991}. The exchange-correlation energy in PBE version of generalized gradient approximation was exploited \cite{perdew}. The accuracy of the real-space numerical integration was specified by the cutoff energy of 200 Ry, the total-energy convergence criterion was $3*10^{-5}$~eV, whereas the surface Brillouin zone of the supercell was sampled with a $7\times 7$ mesh of $k$ points. 

The basis functions were taken as follows: Bi$8.0-s3p3d2f2$, Te$7.0-s3p3d2f1$, Mn$6.0-s3p3d2$, this means that 3 primitive orbitals for each $s$ and $p$, 2 orbitals for $d$ and 1 for $f$ channels with the cutoff radius of 8.0 a.u.. were set to represent the basis functions of bismuth atoms etc. The Mn $3d$ states were treated within the DFT+U approach \cite{Han2006} within the Dudarev scheme \cite{Dudarev1998} where U parameter equals 5 eV \cite{Otrokov2019, Otrokov.prl2019, Eremeev.jac2017, Otrokov.2dmat2017, Otrokov.jetpl2017, Hirahara.ncomms2020, Eremeev.nl2018}.

The surface was represented by a repetitive slab of 6 SLs of MnBi$_2$Te$_4$. The vacuum layer of 12 \AA~ was placed between slabs to avoid their interaction. Effective screening medium method with so-called "vacuum/metal" boundary condition \cite{Otani2006} was used to take into account the impact of an additional surface charge. The spin magnetic moments of the atoms were calculated from the Mulliken population difference.

\section*{Acknowledgement}

The authors acknowledge support by the Saint Petersburg State University Grant No. ID 73028074, Russian Science Foundation Grant No. 18-12-00062 in part of the photoemission measurements and total analysis of the results, Grant No. 18-12-00169 in part of the electronic band structure calculations and Grant No. 20-42-08002 in part of analysis of magnetic properties and Science Development Foundation under the President of the Republic of Azerbaijan Grant No. EI F-BGM-4-RFTF1/2017-21/04/1-M-02. M.M.O. acknowledges the support by  Spanish Ministerio de Ciencia e Innovaci\'on (Grant no. PID2019-103910GB-I00).

%\bibliography{biblio_ME_in_MBT_mo}

%

\end{document}